# Optical conductivity signatures of strong correlations and multiband superconductivity in infinite-layer nickelates

Woo Jin Kim[1,2,3*], Kyuho Lee[1,4], Eun Kyo Ko[1,2], Jaeseok Son[5,6], Yonghun Lee[1,2], Yijun Yu[1,2,7], Soon Jae Moon[8], Tae Won Noh[5,6], and Harold Y. Hwang[1,2*]

[1] Stanford Institute for Materials and Energy Sciences, SLAC National Accelerator Laboratory, Menlo Park, CA 94025, United States.

[2] Department of Applied Physics, Stanford University, Stanford, CA 94305, United States.

[3] Department of Material Science Engineering, Pusan National University, Busan 46241, Republic of Korea.

[4] Department of Physics, Stanford University, Stanford, CA 94305, United States.

[5] Center for Correlated Electron Systems, Institute for Basic Science, Seoul 08826, Republic of Korea.

[6] Department of Physics and Astronomy, Seoul National University, Seoul 08826, Republic of Korea.

[7] State Key Laboratory of Surface Physics and Department of Physics, Fudan University, Shanghai 200438, China

[8] Department of Physics, Hanyang University, Seoul 04763, Republic of Korea.

[*] Correspondence to: woojin@pusan.ac.kr, hyhwang@stanford.edu

## Abstract

Since the discovery of superconductivity in infinite-layer nickelates, there have been extensive efforts to unravel their electronic structure and pairing mechanism. In particular, understanding how the electronic structure evolves with doping is essential for clarifying theoretical models of superconductivity in nickelates. Here we present studies of the optical conductivity of $Nd_{1-x}Sr_xNiO_2$ thin films spanning the full phase diagram $0.025 \leq x \leq 0.30$ using spectroscopic ellipsometry. The data are consistent with a two-band Drude model, which allows the decomposition of the intraband response into distinct contributions. One is from a 'narrow' Drude term which we associate with electron bands, and the other a 'broad' Drude term linked to the hole band with strong correlations. Increasing Sr doping leads to an expansion of the hole band spectral weight, and a corresponding reduction in the electron band, indicative of the multiband electronic structure and a doping-dependent reconstruction of the Fermi surface. Both doping and temperature-dependent optical spectra display significant spectral weight transfer from high to low energy, a hallmark of strong electronic correlations. In the superconducting state at optimal doping ($x = 0.15$), both electron and hole bands contribute to the superconducting condensate, signifying multiband superconductivity.

The discovery of superconductivity in infinite-layer nickelates[1] has attracted intense interest due to its material similarities with high-$T_c$ cuprate superconductors. Infinite-layer nickelates share many features with cuprates, including a similar crystal structure composed of two-dimensional $NiO_2$ planes, analogous to the $CuO_2$ planes in cuprates such as $CaCuO_2$ (ref.[2]). Moreover, both systems have the same nominal $d$-electron filling state ($\sim 3d^9$) (ref.[3]). Despite these similarities, important distinctions in electronic structure and correlation effects have also been observed or proposed. While the cuprates are in the charge-transfer regime, infinite-layer nickelates are closer to the Mott-Hubbard regime because of the large charge transfer gap between Ni $3d$ and O $2p$ (ref.[4,5]). Of particular importance are the multiband characteristics of the infinite-layer nickelates, especially the coexistence of a hole band with Ni $3d_{x2-y2}$ character, and electron bands arising from rare-earth $5d$ orbitals, which lead to self-doping effects and the absence of a Mott insulating parent state[5–11].

This multiband character can facilitate inter-orbital correlations, in addition to the on-site Coulomb interaction. An important current debate is the relevance of these additional interactions for the properties of the normal and superconducting phases. Some theoretical pictures emphasize the correlated cuprate-like Ni $3d_{x2-y2}$ band, and treat the additional weakly correlated electron bands as a perturbation[6,8,12–17]. Others propose a central role for multi-orbital physics, such as the role of Hund's coupling[18–20] or Kondo-like hybridization between the correlated and uncorrelated bands[21–24]. In addition to these qualitatively different perspectives, there is also an important question regarding the level of band filling in the two channels as a function of doping across the superconducting dome. Some calculations indicate the full depletion of the electron band before the onset of superconductivity[7,8], others place the crossover near the peak of the superconducting dome[6,15], while still others predict a multiband state persisting well into the overdoped regime beyond the dome[16,24–26]. One central aim of this work is to experimentally probe band filling as a function of doping and the contributions of the occupied bands to superconductivity.

The other focus of our studies is to determine how carrier doping reshapes the electronic structure of the nickelates. In cuprates, doping-dependent studies have revealed profound changes in spectral weight distribution, coherence, and Fermi surface topology. While DC transport and phase diagrams for nickelates have been extensively reported[9,10,27], insights into the evolution of the electronic structure and the corresponding energy scales can only be gained by broadband spectroscopy. First optical reflectivity measurements have been made on $Nd_{0.8}Sr_{0.2}NiO_2$ film on $SrTiO_3$ (ref.[28]), but the dominant contribution of optical phonons of the substrate presents a significant challenge. Recently, impressive advances in *in situ* sample reduction have enabled angle-resolved photoemission spectroscopy (ARPES), allowing direct visualization of band structures in infinite-layer nickelate thin films, undoped and optimally doped infinite-layer thin films[11,29,30].

However, systematic doping-dependent studies for optical conductivity or ARPES remain challenging. In optical measurements, strong phonon contributions from the $SrTiO_3$ substrate hinder the reliable extraction of the intrinsic optical response of infinite-layer nickelates[28]. While ARPES has provided valuable insights into their electronic structure, questions remain regarding the polar nature of the $NdNiO_2/SrTiO_3$ surface[31,32] and its influence on the measured spectra. Here, $Nd_{1-x}Sr_xNiO_2$ infinite-layer thin films were grown on $(LaAlO_3)_{0.3}(Sr_2TaAlO_6)_{0.7}$ (LSAT) substrates (see Methods). The LSAT substrate is optically transparent over a wide energy range and provides enhanced epitaxial stability[27], allowing us to probe the intrinsic electronic response of the nickelate films. Measuring the optical conductivity across the hole-doped nickelate phase diagram offers a complementary window into their charge dynamics, bridging DC transport and electronic band structure. Optical spectroscopy directly probes the frequency-resolved response, allowing us to trace how spectral weight redistributes between high- and low-energy excitations with doping. These insights are essential for assessing the roles of strong correlations and multiband effects, and for clarifying both the parallels and distinctions between the nickelates and the cuprates.

**Doping-dependent optical conductivity of Nd$_{1-x}$Sr$_x$NiO$_2$**

Here we measured the optical conductivity of Nd$_{1-x}$Sr$_x$NiO$_2$ (NSNO) infinite-layer nickelate thin films for $0.025 \leq x \leq 0.30$ using spectroscopic ellipsometry (see Methods). Figures 1a and 1b display the optical conductivity $\sigma_1(\omega)$ for NSNO films $0.025 \leq x \leq 0.2125$ and $0.2125 \leq x \leq 0.30$, respectively. While ellipsometry in principle provides access to both in-plane and out-of-plane (c-axis) responses, in the present ultrathin films (~5 nm) the signal is dominated by the in-plane component, enabling a precise determination of the *ab*-plane optical conductivity only. The optical conductivity of NSNO over the entire range of composition from $x = 0.025$ to 0.30, shows pronounced Drude peaks. Multiple interband transitions are observed at ~0.4, ~1.4, ~2.5, ~3.5 and ~4.0 eV (Fig. 1a). Among these, the transitions near ~3.5 eV and ~4.0 eV have been assigned to Ni 3*d* orbital to La 4*f* orbital and O 2*p* to La 4*f* orbital excitations, respectively[24] (Extended Data Fig. 1). Notably, since the O 2*p* band lies at a lower energy than the Ni 3*d* band, these results suggest that the on-site Coulomb repulsion ($U$) is smaller than the charge transfer energy ($\Delta$) in NSNO, consistent with observations in the literature[4,5].

Interestingly, we observed a pronounced spectral weight redistribution in the optical conductivity over a broad energy region upon doping (Fig. 1a and 1b). Specifically, Fig. 1a and 1b shows a decrease in conductivity above ~2.5 eV and a corresponding increase below ~2.5 eV, forming an isosbestic point at ~2.5 eV. The presence of the isosbestic point clearly indicates that the spectral weight lost above ~2.5 eV is shifted to lower energies. This behavior closely resembles that of the cuprates (e.g., La$_{1-x}$Sr$_x$CuO$_4$, Nd$_{2-x}$Ce$_x$CuO$_4$, etc.)[33–35], where the spectral weight associated with charge-transfer excitations moves toward lower-energy excitations as doping increases. Such spectral weight transfer over a broad energy region is a hallmark of strongly correlated electron systems, supporting that NSNO resides in this regime[33,36–38].

To provide a quantitative basis for the optical conductivity data, we estimated the effective electron number per Ni atom, defined as the integrated spectral weight up to a photon energy $\hbar\omega$:

$$N_{eff}^*(\omega) = \frac{2m_0 V}{\pi e^2} \int_0^\omega \sigma(\omega') d\omega', \qquad (1)$$

where $m_0$ is taken as the free-electron mass and $V$ is the unit cell volume of NSNO. $N_{eff}^*(\omega)$ is proportional to the number of electrons involved in the optical excitations up to $\hbar\omega$. $N_{eff}^*(\omega)$ from the conductivity data are shown in Fig. 1c and 1d for various $x$. In every sample, a steep initial rise of $N_{eff}^*(\omega)$ is observed at low photon energy, where the intraband conductivity is dominant. For NSNO films with doping levels between $x = 0.025$ and $0.2125$, $N_{eff}^*(\omega)$ increases with doping (Fig. 1c). This increase becomes less pronounced at higher energy, near 4 eV. The overall trend provides quantitative evidence that doping primarily affects electronic excitations below ~4 eV, confirming a spectral weight transfer from higher-energy states (2.5–4 eV) to lower-energy excitations.

Since ~2.5 eV corresponds approximately to an isosbestic point in $\sigma_1(\omega)$, $N_{eff}^*(2.5 \text{ eV})$ serves as a useful measure of the low-energy spectral weight gained at the expense of higher-energy excitations. As shown in Fig. 1e, $N_{eff}^*(2.5 \text{ eV})$, exhibits an anomalous doping dependence: it increases nearly linearly with $x$ in the underdoped regime and consistently exceeds the nominal doping level (i.e., $N_{eff}^* > x$), suggesting additional contributions associated with the reconstruction of the electronic structure beyond the doped holes. The finite value of $N_{eff}^*$ (2.5 eV) even near $x = 0$ is presumably due to the self-doping effects. In the overdoped regime ($0.2125 < x \leq 0.3$), the spectral weight transfer from high to low photon energies becomes suppressed. The increasing trend in $N_{eff}^*$ persists up to $x = 0.2125$ (Fig. 1e), after which it saturates (Fig. 1d). Such doping-dependent behavior is reminiscent of cuprates, where the suppression of charge-transfer excitations in the overdoped regime similarly leads to a saturation or decrease in $N_{eff}^*$ (ref.[33]). Interestingly, the doping level at which this change occurs coincides with the composition where the Hall coefficient of NSNO changes sign (from negative to positive) at temperatures below 80 K[27], suggesting a possible link between the optical spectral weight evolution and changes in the underlying Fermi surface topology.

**Two-band Drude model analysis on Nd$_{1-x}$Sr$_x$NiO$_2$**

To understand the doping evolution of the low-energy electronic structure, we performed a two-band Drude model analysis, considering both the structure of the data and the multi-band nature of NSNO (ref.[11]). We also explicitly confirm poor fitting using a single Drude component to demonstrate that two Drude components are necessary (Extended Data Fig. 2). Figures 2a–e shows the low-energy (< 1.5 eV) optical conductivity for different Sr doping levels $x$, along with the corresponding Drude–Lorentz fits:

$$\sigma_1(\omega) = \frac{1}{4\pi}\left[\sum_i \frac{\omega_{p,i}^2 \tau_i}{(\omega^2 \tau_i^2 + 1)} + \sum_j \frac{\gamma_j \omega^2 \omega_{p,j}^2}{(\omega_{0,j}^2 - \omega^2)^2 + \gamma_j \omega^2}\right]. \tag{2}$$

The first term refers to the sum of Drude components, which describe the optical response of free carriers, with $\omega_{p,i}$ and $1/\tau_i$ corresponding to the plasma frequency and the scattering rate, respectively. The second term represents a sum of Lorentzian oscillators, which are used to model the interband transitions. In the Lorentz term, $\omega_{p,j}$, $\gamma_j$ and $\omega_{0,j}$ are the resonance frequency, damping, and plasma frequency of the $j^{\text{th}}$ excitation, respectively. Note that below 1.5 eV, the optical conductivity consists of two Drude components and two interband transition peaks, located at ~0.4 eV and ~1.4 eV, respectively.

Our analysis of the two Drude components quantifies the multiband nature of the low-energy electronic response of NSNO[6–8,14,15,39–41], with one narrow and the other broad (green and purple lines in Figs. 2a–e). With increasing Sr doping, the spectral weight of the narrow Drude peak decreases, while that of the broad component increases. Note that the extracted $\omega_p^2$ from the Drude fitting is proportional to the spectral weight of the corresponding Drude component. Since $\omega_p^2 = 4\pi n e^2 / m^*$, where $n$ is the carrier density, the decrease of the narrow component and increase of the broad one (Fig. 2f) reflect a redistribution of carriers between the two bands with doping (Extended Data Fig. 3). The broad (narrow) Drude component is attributed to the intraband excitations of the hole (electron) band. This two-band picture is corroborated by recent electronic structure calculations[6–8,14,15,39–41], DC transport[9,10,27], and ARPES measurements[11,29]. Together, these studies point towards a consensus on the electronic structure for NdNiO$_2$: a large Ni $3d_{x2-y2}$ hole pocket coexisting with two electron pockets of Nd $5d_{xy}$ and $5d_{3z2-r2}$. As Sr doping is introduced, the hole pocket expands while the electron pockets diminish (Fig. 2h).

Having assigned the broad and narrow Drude components to the hole and electron bands, respectively, we discuss the doping evolution of their scattering rates $1/\tau$ extracted from the Drude fits. As shown in Fig. 2g, the scattering rate for the hole band gradually decreases with Sr doping up to $x = 0.20$. We note that the increase in $1/\tau$ for the hole band in the overdoped regime ($0.2125 < x \leq 0.3$) may be associated with an incoherent state induced by disorder arising at the limits of high doping[27]. In contrast, the scattering rate for the electron band shows a much smaller decrease across the entire doping range. The decrease in the hole band's scattering rate $1/\tau$ (Fig. 2g), and the accompanying redistribution of spectral weight toward the Drude peak (Fig. 2f), indicate a doping-induced enhancement of coherent quasiparticle states. This behavior reflects a correlation-driven crossover from an incoherent, Mott-like regime to a coherent metallic state.

**Temperature-dependent optical conductivity of optimally doped $Nd_{0.85}Sr_{0.15}NiO_2$**

To investigate the charge dynamics of both the electron and hole bands in superconducting NSNO, we examined the temperature-dependent optical conductivity of the optimally doped sample ($x = 0.15$). This sample enters the superconducting phase at $T_{c,onset} = 22$ K (Extended Data Fig. 4). The optical conductivity at all measured temperatures is well described by two Drude components (Figs. 3a–e), and the fitting results agree with the DC conductivity values (red square in Fig. 3a, b, and f). In the normal state, $\omega_p^2$ for the hole (electron) band increases (decreases) as the temperature decreases, indicating that the Fermi surface of the hole (electron) band expands (shrinks) upon cooling (Fig. 3h). This trend is consistent with the temperature dependence of the Hall coefficient: $R_H$ becomes less negative at lower temperatures[1,9,10,27], implying a relative increase (decrease) in the hole (electron) contribution to transport.

To examine the temperature-dependent evolution of the electronic structure in the normal state, we analyzed the effective electron number ratio, $N_{eff}^*(150$ K$)/ N_{eff}^*(300$ K$)$, as shown in Fig. 3g. At all photon energies, this ratio exceeds unity, indicating a spectral weight transfer from higher to lower energy. This behavior is a signature of strong correlations, where electronic coherence gradually develops as the temperature decreases. In Mott–Hubbard systems, such spectral weight redistribution manifests as an enhancement of the Drude weight, often interpreted as increased carrier kinetic energy upon cooling[42–44]. These results reinforce the view that NSNO resides in a strongly correlated regime, akin to doped cuprates and other correlated oxides.

On the other hand, when the temperature falls below $T_{c,onset}$, the effective electron number ratio, e.g., $N_{eff}^*(15 \text{ K})/N_{eff}^*(150 \text{ K})$ begins to decrease (blue solid line in Fig. 3g), signaling the formation of a superconducting condensate. Notably, our analysis reveals that both the hole and electron bands contribute to the superconducting condensate, meaning $\omega_p^2$ for both hole and electron bands is reduced upon cooling below $T_{c,onset}$ (Fig. 3h). The sharp drop in $1/\tau$ for both bands below $T_{c,onset}$ signals the loss of normal-state carriers as they condense into the superfluid (Fig. 3i). This multiband involvement in the superconducting phase aligns with the normal-state behavior discussed earlier, where two distinct Drude components were required to describe the conductivity. The participation of both carrier types in the superconducting condensate suggests that superconductivity in NSNO is multiband in nature.

**Discussion**

With Sr doping and decreasing temperature, the coherent (Drude) peak for NSNO is enhanced. This is accompanied by a compensating loss of spectral weight at higher energies, demonstrating a correlation-driven redistribution of charge dynamics in NSNO. Such behavior mirrors a hallmark of the cuprates, where coherent quasiparticles emerge from an incoherent background, but in the infinite-layer nickelates, it occurs within a Mott–Hubbard framework rather than a charge-transfer regime. Our two-band Drude analysis further reveals contributions from both Ni $3d$-derived hole carriers and Nd $5d$-derived electron carriers[11,29]. As the Sr doping level increases, the hole pocket expands while the electron pockets shrink, yet our results show no evidence of complete electron depletion even beyond the superconducting dome. The suppression of both Drude components below $T_{c,onset}$ indicates that carriers from both bands condense into the superconducting state, pointing to a correlated multiband superconducting phase. Importantly, this behavior sets the nickelates apart from the predominantly single-band nature of cuprates, suggesting that the additional electron pockets in the infinite-layer nickelate fermiology should not be treated as mere bystanders but instead important contributors to the establishment of superconductivity in this material system. Taken together, these results suggest that infinite-layer nickelates combine strong electronic correlations with multiband physics and may thus be compatible with recently proposed multiband pairing scenarios[7,22].

# Main figure legends

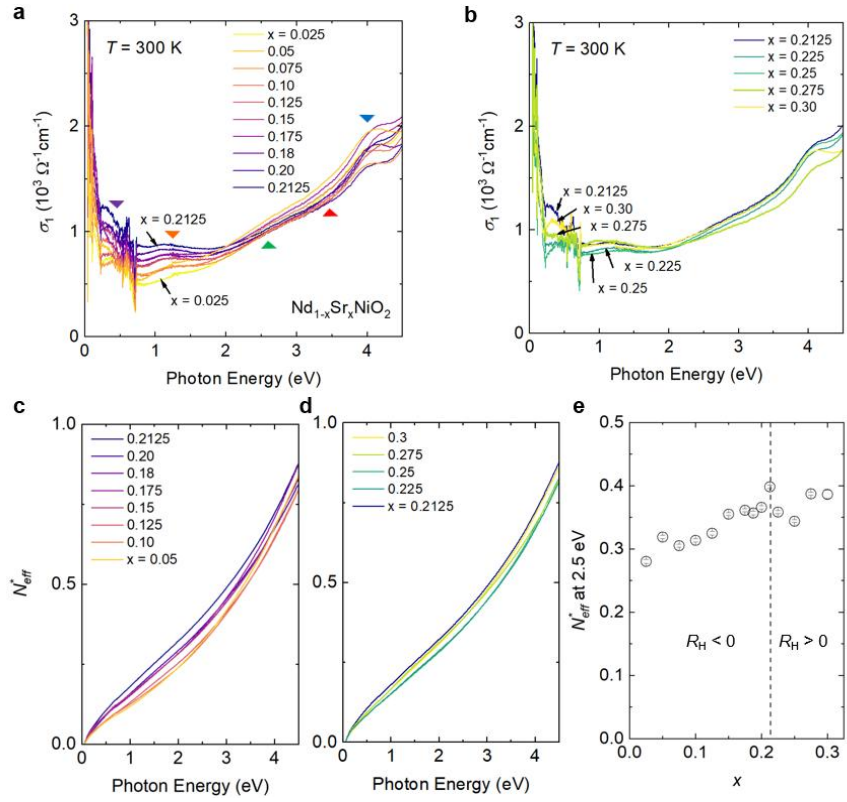

**Figure 1 | Doping-dependent optical conductivity of Nd$_{1-x}$Sr$_x$NiO$_2$.** Optical conductivity $\sigma_1(\omega)$ obtained from spectroscopic ellipsometry with $E//ab$: **a**, $0.025 \leq x \leq 0.2125$ and **b**, $0.2125 \leq x \leq 0.30$. Effective electron number per Ni atom, $N_{\mathrm{eff}}^{*}(\omega)$: **c**, $0.025 \leq x \leq 0.2125$ and **d**, $0.2125 \leq x \leq 0.30$. **e**, $N_{\mathrm{eff}}^{*}$ at 2.5 eV as a function of Sr composition $x$.

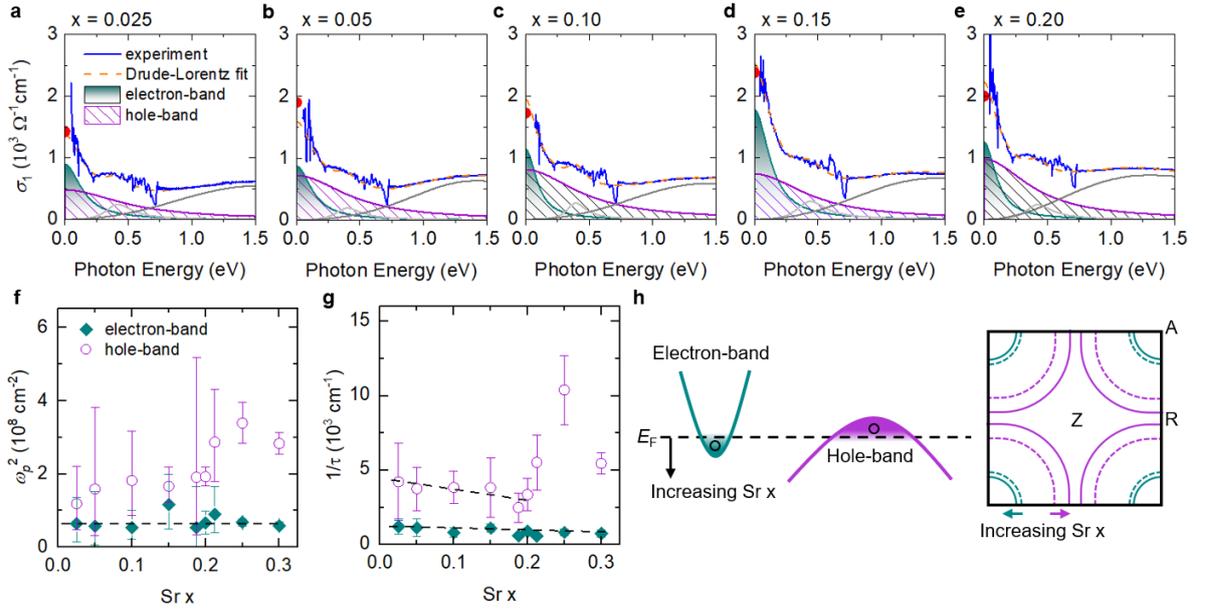

**Figure 2 | Two-band Drude model analysis of Nd$_{1-x}$Sr$_x$NiO$_2$.** Low-energy (< 1.5 eV) optical conductivity at $T$ = 300 K for different Sr doping levels: **a**, $x$ = 0.025, **b**, $x$ = 0.05, **c**, $x$ = 0.10, **d**, $x$ = 0.1875, **e**, $x$ = 0.20. Red circles indicate the DC conductivity, solid blue and dashed lines represent the experimental $\sigma_1(\omega)$ and Drude–Lorentz fit results, respectively. Green and purple lines correspond to the narrow and broad Drude components, respectively. Two Lorentzian contributions from the Drude-Lorentz fit are shown as light gray and gray lines. **f**, The Drude spectral weight $\omega_p^2$ and **g**, the scattering rate $1/\tau$ extracted through the two-band Drude model. **h**, A schematic illustration of the two-band electronic structure showing Fermi-level shifts with Sr doping assuming rigid bands.

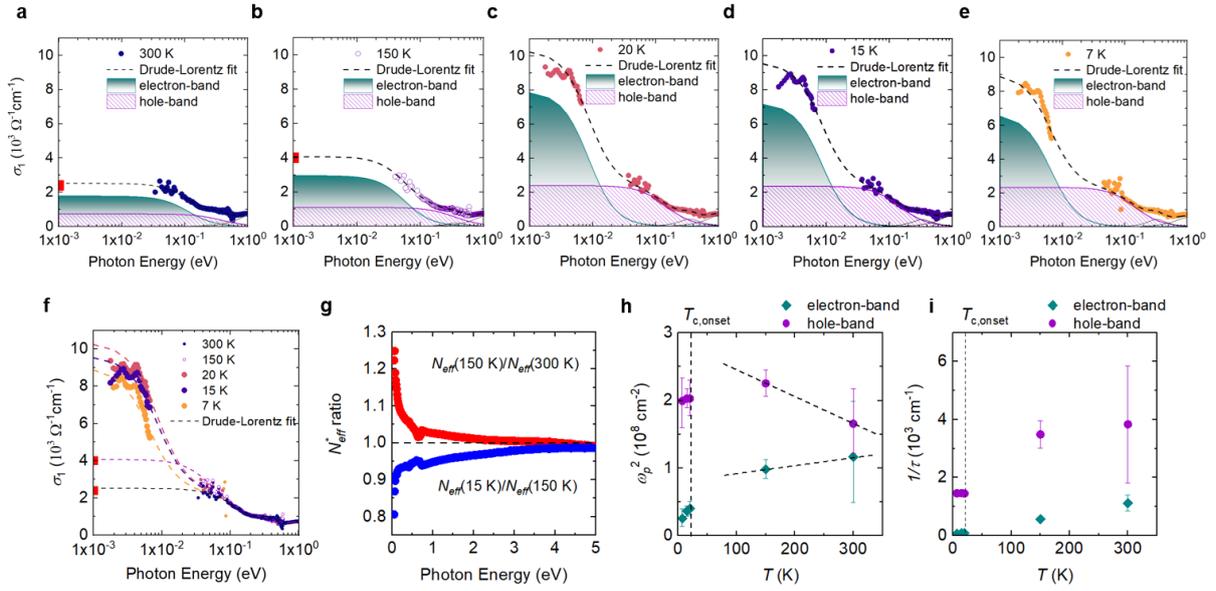

**Figure 3 | Temperature-dependent in-plane optical conductivity of $Nd_{0.85}Sr_{0.15}NiO_2$.** Two-band Drude model fits to low-energy (< 1.2 eV) optical conductivity $\sigma_1(\omega)$ at various temperatures: **a**, $T = 300$ K, **b**, $T = 150$ K, **c**, $T = 20$ K, **d**, $T = 15$ K, **e**, $T = 7$ K. Red squares indicate the DC conductivity, and dashed lines represent the Drude–Lorentz fit results. Green and purple lines correspond to the narrow and broad Drude components, respectively. **f**, combined $\sigma_1(\omega)$ spectra from panels **a-e**. **g**, Effective electron number ratio is plotted at different temperatures: red circles show the ratio of $N_{eff}^*$ at 150 K to that at 300 K ($N_{eff}^*(150$ K$)/N_{eff}^*(300$ K$)$), while blue circles represent the ratio of $N_{eff}^*$ at 15 K to 150 K ($N_{eff}^*(15$ K$)/N_{eff}^*(150$ K$)$). **h**, The Drude spectral weight ($\omega_p^2$) and **i**, the scattering rate ($1/\tau$) derived by the two-band Drude model fit across the temperatures. Note that the low energy spectra below 10 meV is reproduced from [ref.[45]]. [Reproduced with permission from Cheng *et al.*, Nature Materials 23, 775–781 (2024). Copyright 2024 Nature Materials.]

## Methods

### Synthesis of Nd$_{1-x}$Sr$_x$NiO$_2$ thin films

Perovskite Nd$_{1-x}$Sr$_x$NiO$_3$ thin films, approximately 15 unit cells (u.c.), were synthesized by pulsed laser deposition (PLD) on perovskite (LaAlO$_3$)$_{0.3}$(Sr$_2$TaAlO$_6$)$_{0.7}$ (001) substrates. These substrates are known to optimize the epitaxial mismatch for both the perovskite precursor and infinite-layer phases enabling the growth of high-quality infinite-layer thin films[27]. We used a Nd$_{1-x}$Sr$_x$NiO$_3$ polycrystalline target and a KrF excimer laser ($\lambda = 248$ nm, pulse repetition rate of 8 Hz, and laser fluence of 2.6 J/cm$^2$). The distance between the target and the substrate was maintained at 55.5 mm. The optimal conditions for high-quality Nd$_{1-x}$Sr$_x$NiO$_3$ films were found to be a substrate temperature $T = 580$ °C under oxygen partial pressure $P_{O2} = 150$ mTorr. To protect the nickelate films from potential degradation during and after the reduction process, we capped the as-grown Nd$_{1-x}$Sr$_x$NiO$_3$ with 4 u.c. of SrTiO$_3$. The precursor Nd$_{1-x}$Sr$_x$NiO$_3$ film was then reduced to Nd$_{1-x}$Sr$_x$NiO$_2$ by topotactic reduction. We loosely covered the Nd$_{1-x}$Sr$_x$NiO$_3$ film with aluminum foil and sealed it with ~ 0.1 g of CaH$_2$ powder in a vacuum glass tube. We annealed the sealed tube at 240-260 °C for ~2.5-4 hours in a tube furnace, with ramp rate of 10 °C/min.

### Spectroscopic ellipsometry measurements

The ellipsometric measurements were performed using a spectroscopic ellipsometer (J. A. Woollam Inc.) in the energy range 0.05 to 6.5 eV. The ellipsometric parameters $\Psi$ and $\Delta$ are defined by tan($\Psi e^{i\Delta}$) = $r_p/r_s$, where $r_p$ and $r_s$ are the complex Fresnel coefficients for light polarized parallel and perpendicular to the plane of incidence, respectively. From $\Psi(\omega)$ and $\Delta(\omega)$, we directly determined the real and imaginary parts of the complex dielectric function, $\varepsilon(\omega) = \varepsilon_1(\omega) + i\varepsilon_2(\omega)$, and the related optical conductivity $\sigma_1(\omega) = \omega\varepsilon_2(\omega)/4\pi$. The ellipsometric data were analyzed using a point-by-point regression method within a film-on-substrate model. We first measured the bare LSAT substrate and the SrTiO$_3$ (4 u.c.)/LSAT heterostructure. Their optical conductivities were extracted independently and then used to remove the substrate and capping-layer contributions from the total film response, yielding the intrinsic optical conductivity of NSNO.

## Acknowledgments

This work was supported by the Department of Energy (DOE), Office of Basic Energy Sciences, Division of Materials Sciences and Engineering (contract No. DE-AC02-76SF00515). Additional support was provided by Gordon and Betty Moore Foundation's Emergent Phenomena in Quantum Systems Initiative (grant No. GBMF9072, synthesis equipment), and the Kavli Foundation, Klaus Tschira Stiftung, and Kevin Wells (optical analysis). W.J.K. acknowledges funding by the Regional Innovation System & Education (RISE) program through the Institute for Regional Innovation System & Education in Busan Metropolitan City, funded by the Ministry of Education (MOE) and the Busan Metropolitan City, Republic of Korea (2025-RISE-02-004-14300001-01) and National Research Foundation (NRF) of Korea (No. RS-2024-00404737). S.J.M. was supported by the National Research Foundation of Korea (NRF) grant funded by the Korea government (MSIT) (No. RS-2024-00416036).

## Author contributions

W.J.K. and H.Y.H. conceived and designed the experiments. W.J.K., E.K.K., J.S., and T.W.N performed spectroscopic ellipsometry measurements and analysis. K.L. and Y.L. grew and characterized the samples. W.J.K., Y.Y., S.J.M. and H.Y.H. wrote the manuscript, with input from all authors.

## Data availability

The data presented in the figures and other findings of this study are available from the corresponding authors upon reasonable request.

## Competing financial interests

The authors declare no competing financial interests.

**Extended data legends**

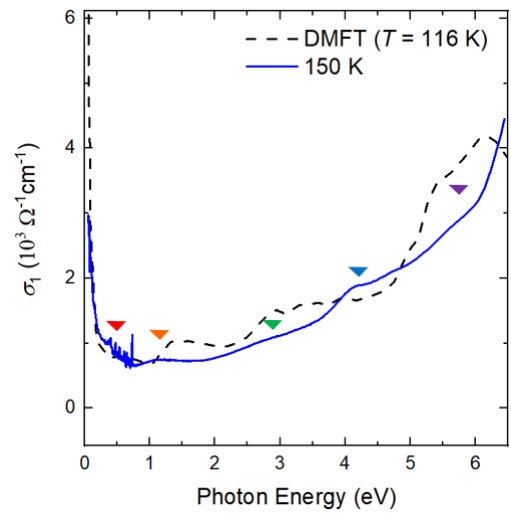

**Extended Data Fig. 1 | Comparison of calculated and experimental optical conductivity.** Note that the optical conductivity data calculated using DMFT is reproduced from [ref.24].

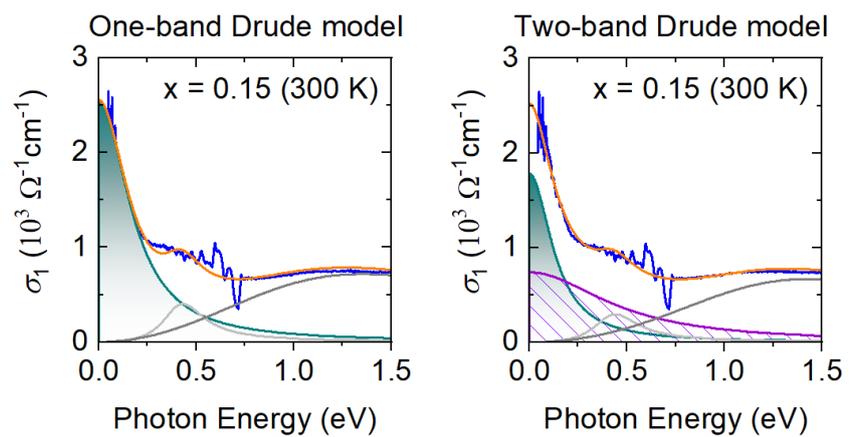

**Extended Data Fig. 2 | Validation of two-band Drude model analysis.**

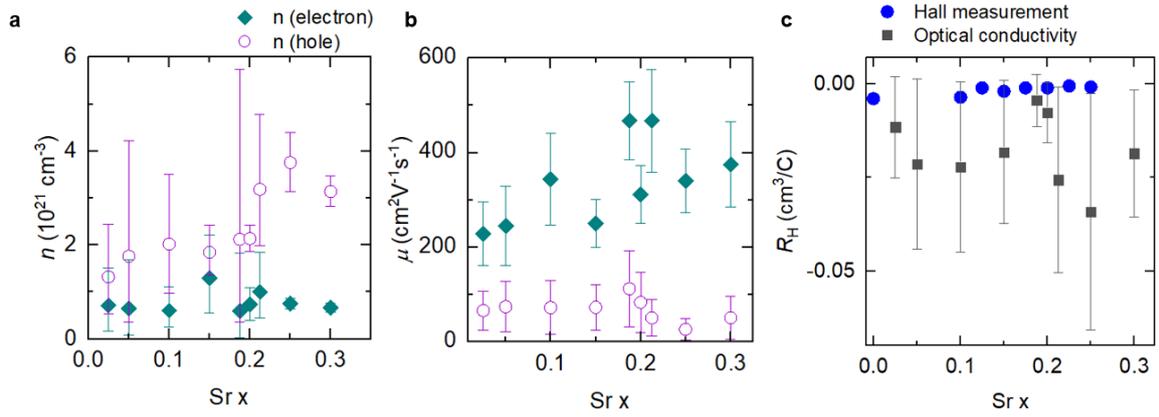

**Extended Data Fig. 3 | Estimation of Hall coefficients of Nd$_{1-x}$Sr$_x$NiO$_2$ from optical conductivity.**
**a**, Carrier densities for both hole and electron bands estimated from the two-Drude analysis. **b**, Corresponding mobility values for the hole and electron carriers. **c**, Hall coefficients calculated from the extracted carrier densities and mobilities. Hall coefficient data obtained from direct Hall measurements are reproduced from [ref.[10]] for comparison.

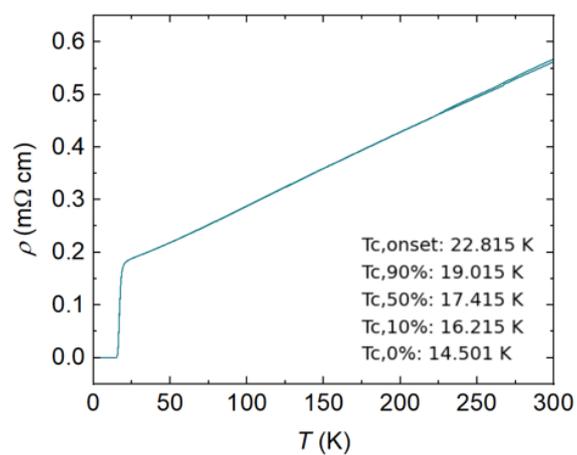

**Extended Data Fig. 4 | Temperature dependent resistivity of thin film Nd$_{0.85}$Sr$_{0.15}$NiO$_2$.**